\documentclass[%
 aip,
 amsmath,amssymb,
reprint,%
]{revtex4-1}

\usepackage{graphicx,subfig}

\usepackage{dcolumn}
\usepackage{bm}

\usepackage[utf8]{inputenc}
\usepackage[T1]{fontenc}
\usepackage{mathptmx}
\usepackage{etoolbox}
\newcommand{\cb}{\color{black}}
\newcommand{\cred}{\color{black}}
\newcommand{\cn}{\color{black}}

\newcommand{\bX}{\boldsymbol{\textbf{X}}}
\newcommand{\bA}{\boldsymbol{\textbf{A}}}
\newcommand{\x}{\boldsymbol{\textbf{x}}}

\newcommand{\bD}{\boldsymbol{\textbf{D}}}

\newcommand{\Hmat}{\boldsymbol{\textbf{H}}}
\newcommand{\Hmatp}{\boldsymbol{\textbf{H}^{p}}}
\newcommand{\smallsim}{\smallsym{\mathrel}{\sim}}
\makeatletter
\newcommand{\smallsym}[2]{#1{\mathpalette\make@small@sym{#2}}}
\newcommand{\make@small@sym}[2]{%
  \vcenter{\hbox{$\m@th\downgrade@style#1#2$}}%
}
\newcommand{\downgrade@style}[1]{%
  \ifx#1\displaystyle\scriptstyle\else
    \ifx#1\textstyle\scriptstyle\else
      \scriptscriptstyle
  \fi\fi
}
\makeatother
\usepackage{xcolor}
\definecolor{red}{rgb}{1,0,0}
\makeatletter
\def\@email#1#2{%
 \endgroup
 \patchcmd{\titleblock@produce}
  {\frontmatter@RRAPformat}
  {\frontmatter@RRAPformat{\produce@RRAP{*#1\href{mailto:#2}{#2}}}\frontmatter@RRAPformat}
  {}{}
}%
\makeatother
\begin{document}

\preprint{AIP/123-QED}

\title{\large Subspace recursive Fermi-operator expansion strategies for large-scale DFT eigenvalue problems on HPC architectures}

\author{Sameer Khadatkar}
\author{Phani Motamarri}
 \email{phanim@iisc.ac.in}
\affiliation{ 
Department of Computational and Data Sciences, Indian Institute of Science, Bengaluru 560012, India.
}%


\begin{abstract}
Quantum mechanical calculations for material modelling using Kohn-Sham density functional theory (DFT) involve the solution of a nonlinear eigenvalue problem for $N$ smallest eigenvector-eigenvalue pairs with $N$ proportional to the number of electrons in the material system. These calculations are computationally demanding and have asymptotic cubic scaling complexity with the number of electrons. Large-scale matrix eigenvalue problems arising from the discretization of the Kohn-Sham DFT equations employing a systematically convergent basis traditionally rely on iterative orthogonal projection methods, which are shown to be computationally efficient and scalable on massively parallel computing architectures.  However, as the size of the material system increases, these methods are known to incur dominant computational costs through the Rayleigh-Ritz projection step of the discretized Kohn-Sham Hamiltonian matrix and the subsequent subspace diagonalization of the projected matrix. This work explores the potential of polynomial expansion approaches based on recursive Fermi-operator expansion as an alternative to the subspace diagonalization of the projected Hamiltonian matrix to reduce the computational cost. Subsequently, we perform a detailed comparison of various recursive polynomial expansion approaches to the traditional approach of explicit diagonalization on both multi-node CPU and GPU architectures and assess their relative performance in terms of accuracy, computational efficiency, scaling behaviour and energy efficiency.
\end{abstract}

\maketitle



\section{\textbf{Introduction}}\label{sec:level1}
\vspace{-0.2in}
A well-known and challenging application of eigenvalue problems is in the area of quantum modelling of materials using Kohn-Sham density functional theory (DFT)~\cite{PhysRev.140.A1133}, which has been immensely successful in providing critical insights into various ground-state material properties. To compute the ground-state electronic structure in DFT, one is confronted with solving a large-scale nonlinear eigenvalue problem using a self-consistent field iteration procedure (SCF) for $N$ smallest eigenvalue/eigenvector pairs, with $N$ being proportional to the number of electrons in the material system. This results in asymptotic cubic complexity $\mathcal{O}(N^3)$ with the number of electrons for DFT, making these calculations computationally demanding and often restrictive in terms of system sizes that can be handled using widely used DFT codes. Many of these codes employ systematically convergent plane-wave basis sets restricting the simulation domains to be periodic, or employ atomic-orbital (AO) type basis sets with very few basis functions per atom, but AO basis sets are not complete. Furthermore, the extended nature of these basis sets restricts their parallel scalability. To extend the range of system sizes to be studied, numerous past efforts have focused on developing systematically convergent real-space computational methodologies ~\cite{wavelet2020,phanish2017,46PFLOP,motam2020,das2022, motam2013, motam2014} that have relied on reducing the prefactor associated with the cubic computational complexity alongside improving the parallel scalability, thereby enabling large-scale DFT calculations up to 100,000 electrons. These real-space DFT discretization approaches result in large sparse Hermitian eigenvalue problems to be solved for $N$ smallest eigenvalue/eigenvector pairs.

DFT codes employing systematically convergent basis sets either in Fourier-space or in real-space require large number of basis functions per atom to achieve chemical accuracy compared to approaches employing AO basis sets. This results in large Hamiltonian matrices of dimension $M$ with $M \approx 10^{5} - 10^{8}$, whose $N$ smallest eigenvalue/eigenvector pairs we seek. We note that $N$ is usually $0.1-0.5\%$ of $M$, the total number of basis functions (degrees of freedom (DoFs)) used in the simulation. The most popular eigensolver strategies employed to solve these large-scale DFT eigenvalue problems include the Davidson approach~\cite{DAVIDSON197587}, Locally Optimal Block Pre-conditioned Conjugate Gradient (LOBPCG) method~\cite{lobpcg}, or the Chebyshev filtered subspace iteration (ChFSI)~\cite{chfsi, motam2013} approach. These eigensolvers belong to the category of iterative orthogonal projection methods (IOP) wherein the discretized Kohn-Sham Hamiltonian matrix $\Hmat$ of size $M \times M$ is orthogonally projected onto a carefully constructed subspace rich in the wanted eigenvectors (Rayleigh-Ritz step), and subsequently, the resulting smaller dense projected Hamiltonian $\Hmatp$ of dimension $k (<< M)$ is explicitly diagonalized (subspace diagonalization) to approximate the desired eigenvalue/eigenvector pairs of the $\Hmat$ matrix. The cubic scaling computational cost of this subspace diagonalization step dominates for medium to large-scale material systems ($N > 20,000$) compared to the costs associated with subspace construction and Rayleigh-Ritz steps in IOP methods. For instance, Das et al.\cite{das2022} employing the ChFSI approach have reported that the subspace diagonalization constitutes roughly 30\% of the total ChFSI cost for $N \approx 30,000$, whereas it accounts for around 56\% of the total cost for $N \approx 50,000$. To this end, the current work explores recursive Fermi-operator expansion approaches on HPC architectures as an alternative to the subspace diagonalization procedure to improve the computational efficiency, thereby reducing the computational prefactor associated with the cubic complexity of the DFT problem. Furthermore, the accuracy, computational efficiency,  parallel performance and energy efficiency of these approaches are examined on both multi-node central processing unit (CPU) and graphics processing unit (GPU) architectures and compared with explicit diagonalization of the projected Hamiltonian matrix.

Recursive polynomial expansion approaches (RPE) rely on the key idea that constructing a density matrix (projector matrix corresponding to $N$ smallest eigenvectors) suffices to compute ground-state electronic structure in DFT at zero-temperature without the necessity of computing explicit eigenvalues and eigenvectors. In the past, RPE approaches \cite{RFOE3,ARFOE3,TRFOE4,TRFOE2} have been used for ground-state DFT calculations with atomic-orbital basis. Most of these methods aimed to achieve linear scaling complexity using sparse-matrix algebra employing numerical thresholds. Further, these recursive expansion schemes have also been shown to achieve high throughput performance using dense matrix algebra\cite{gpurfoe2012} on a single Nvidia M2070 GPU, compared to traditional diagonalisation approaches. More recently, the computational structure of RPE approaches has been leveraged in the form of a deep neural network architecture\cite{Fink2021April} to take advantage of tensor cores in Nvidia A100 GPUs. These techniques have been applied for \emph{ab initio} molecular dynamics \cite{Fink2021Oct} and quantum response calculations \cite{Fink2022} using atomic-orbital basis sets. Furthermore, Google's tensor processing units (TPUs) have been successfully used for large-scale DFT calculations \cite{PedersonTPU} employing density matrix purification approaches, which are similar in spirit to RPE and involve dense matrix-matrix multiplications\cn. However, to the best of our knowledge, computational efficiency, scaling behaviour on distributed systems and energy efficiency of these RPE approaches have not been explored compared to subspace diagonalization procedures for their use in iterative orthogonal projection methods on multi-node CPU and GPU architectures. The evolving computing architectures in today's exascale era demand scalable methodologies focusing on reduced data movement and increased arithmetic intensity on massively parallel computing architectures, with an equal emphasis on employing energy-efficient algorithms. The current work exploring recursive polynomial expansion approaches as an alternative to subspace diagonalization is towards this direction, enabling large-scale eigenvalue problems arising from the discretization of DFT problem using systematically convergent basis sets employing IOP methods. To this end, the key contributions of our current work, as described in the subsequent sections, include -- (a) efficient implementation strategies of various recursive polynomial expansion (RPE) techniques based on Fermi-operator expansion on both multi-node CPU and GPU architectures for both zero-temperature case, and the finite-temperature case of Fermi-Dirac smearing of the occupancy function, (b) mixed precision strategies in conjunction with RPE to reduce compute and data access costs, (c) assessing accuracy, computational efficiency, parallel scaling behaviour and energy efficiency of the proposed implementation procedures by comparing  with explicit diagonalization algorithms provided by state-of-the-art ELPA library \cite{ELPA, ELPA2}.  
\vspace{-0.3in}
\section{\textbf{Related Work and Background}}\label{sec:level2}
\vspace{-0.1in}
This section discusses key ideas central to recursive polynomial expansion approaches among the various purification schemes \cite{RevModPhys.32.335,PhysRevB.58.12704,RFOE1,RFOE2,RFOE4,RFOE5,RFOE6,RFOE7,RFOE8ARFOE1,BOWLER199995,RFOE9ARFOE2} employed in the literature to approximate the density matrix.
\vspace{-0.2in}
\subsection{Density matrix}
\vspace{-0.1in}
At zero electronic temperature, the density matrix ($\bD$) can be defined as a projector matrix corresponding to the lowest occupied ($N_{occ} \leq N$) eigenvectors of the discretized Kohn-Sham Hamiltonian matrix $\Hmat$. Mathematically, it takes the form of a shifted Heaviside step function, $\theta(.)$, given by
$\bD = \theta[\mu\textbf{I}-{\Hmat}]$. The density matrix ($\bD$) in the case of finite-temperature is a smeared version of zero-temperature density matrix and mathematically represented by a Fermi-operator matrix function given by $ \bD = [e^{\beta({\Hmat}-\mu\boldsymbol{\textbf{I}})}+\boldsymbol{\textbf{I}}]^{-1}$, where, $\boldsymbol{\textbf{I}}$ denotes the identity matrix, $\beta$ = $1/(k_BT_e)$ is the inverse electronic temperature, $\mu$ is the Fermi-energy. Note that the eigenvalues $f_i$ of $\bD$ are referred to as occupancies. $f_i$ is either 0 or 1 for a zero-temperature case, whereas for the case of a finite-temperature case, $f_i \in [0,1]$.

\vspace{-0.2in}
\subsection{\label{sec:level3}Recursive polynomial expansion techniques to approximate the density matrix}
\vspace{-0.1in}
Two types of polynomial expansion schemes can be used to approximate the density matrix -- {\cred(a) Serial Fermi-operator expansion schemes (Chebyshev Fermi-operator expansion scheme \cite{RFOE_ORIG,RevModPhys.78.275}, Green’s function expansion scheme \cite{ZELLER1982993, PhysRevB.75.035123}), (b) Recursive Fermi-operator expansion schemes\cite{RFOE3,ARFOE3,TRFOE4,TRFOE2}}. \cb In this work, we employ the recursive Fermi-operator expansion schemes as they are shown to reach a very high polynomial order in the approximation with few iterations (few matrix-matrix multiplications) \cn and can be used to approximate the density matrix for both zero-temperature, and finite-temperature cases as well.
\vspace{-0.1in}
\subsubsection{\label{sec:level4}Recursive Fermi-operator expansion for zero-temperature density matrix (Z-RFOE)}
\vspace{-0.1in}
The recursive Fermi-operator expansion\cite{RFOE3} involves successive projections of a matrix $\bX_n$, that begins with $\bX_0={\Hmat}$ and $\bX_{n+1}=F_n(\bX_n)$. The functions $F_n(\bX_n)$ are chosen to project the eigenvalue spectrum of $\bX_n$ to eigenvalues closer either to 1 or to 0. Mathematically this can be represented as
\begin{equation}\label{eq:28}
    \bD = \theta(\mu\textbf{I}-{\Hmat}) \approx F_m(F_{m-1}(...F_0(\Hmat)...))
\end{equation}
One of the most efficient techniques in Z-RFOE is to use the {\cred second-order projection polynomials (SP2) ~\cite{ARFOE3}} given by $\bX_{n+1} = F_n(\bX_n) = \bX_n\pm(\bX_n-\bX_n^2)$. The SP2 here is continuously increasing and decreasing function in the interval [0, 1]. The $\pm$ sign is chosen to adjust the trace of $\bX_{n+1}$ in each projection such that it converges to $N_{occ}$.
\vspace{-0.2in}
\subsubsection{\label{sec:level5}Accelerated recursive polynomial expansion for zero-temperature density matrix (A-Z-RFOE)} 
\vspace{-0.1in}
This technique works on the concept of shifting and scaling of the SP2 projection polynomials. In contrast to Z-RFOE which uses fixed SP2 polynomial taking the form $F(\bX)=\bX^2$ or $F(\bX)=2\bX-\bX^2$, the {\cred A-Z-RFOE~\cite{RFOE8ARFOE1, ARFOE4}} approach offers the flexibility to choose the expansion polynomials based on HOMO (Highest Occupied Molecular Orbital) and LUMO (Lowest Unoccupied Molecular Orbital) eigenvalues such that it moves the eigenvalues closer to either 1 or 0 at a rapid rate.
\vspace{-0.2in}
\subsubsection{\label{sec:level6}Recursive Fermi-operator expansion scheme for finite-temperature cases (T-RFOE)}
\vspace{-0.1in}
{ \cred Finite-temperature density matrix has occupancies $f_i \in [0,1]$ and the SP2 recursion scheme discussed above is not well suited for approximating density matrix with fractional occupancies. To this end, an intermediate function generated in Z-RFOE that is obtained before the convergence of the algorithm to the projector matrix is used. This serves as a smeared function to zero-temperature density matrix (Heaviside step function). To this end, the truncated expansion for computing, the density matrix $\bD$ can be given by the expression in Eq. \eqref{eq:211}.
\begin{equation}\label{eq:211}
    G_m(\Hmat) = F_m(F_{m-1}(...F_0(\Hmat)...))
\end{equation}
Lower the electronic temperature $T_e$ higher will be the $\beta$ value (refer to Sec.IIA), and more recursion steps $m$ will be required to approximate the density matrix \cite{TRFOE2, TRFOE4}.}
\vspace{-0.1in}
\subsection{\label{sec:level7}{Accuracy and performance of polynomial expansion procedures}}
\vspace{-0.1in}
The accuracy~\cite{RevModPhys.71.1085} of the aforementioned polynomial expansion procedures can be related to the degree $n_{pl}$ of the polynomial needed to approximate the density matrix and, we note that $n_{pl}\propto (\varepsilon_N-\varepsilon_1)$ with $\varepsilon_1,\varepsilon_N$ being spectral bound estimates of the DFT discretized Hamiltonian $\Hmat$. Since the spectral width of $\Hmat$ arising in DFT calculations employing systematically convergent basis is usually of the $\mathcal{O}(10^{3} - 10^{5})$, a large value of $n_{pl}$ is required for approximating the density matrix. To this end, the above RPE approaches involving $\Hmat$ require large number of computationally expensive matrix-matrix multiplications of huge sizes on distributed architectures and are never done in practice.

\vspace{-0.2in}
\section{\label{sec:level8}Computational methodology and implementation}
\vspace{-0.1in}
\subsection{\label{sec:level9}Methodology and algorithmic details}
\vspace{-0.1in}
As discussed before, we consider the iterative orthogonal projection (IOP) based methods for solving the large-scale eigenvalue problem associated with matrix $\Hmat$ of dimension $M \times M$. In these approaches, the orthogonal projection of the matrix $\Hmat$ is carried out onto a carefully constructed subspace of dimension $k << M$ rich in the wanted eigenvectors. We choose to work with this smaller projected Hamiltonian $\Hmatp$ of dimension $k \times k$ and  employ the recursive polynomial expansion procedures on $\Hmatp$ to approximate the density matrix in the subspace \cite{motam2014} as an alternative to explicit subspace diagonalization of $\Hmatp$. Owing to the reduced dimension of $\Hmatp$ and spectral width of $\Hmatp$ being commensurate with that of occupied Kohn-Sham eigenspectrum, the proposed approach is computationally efficient as demonstrated subsequently.



\begin{figure}[!btp]
    \centering
    \includegraphics[width=\linewidth]{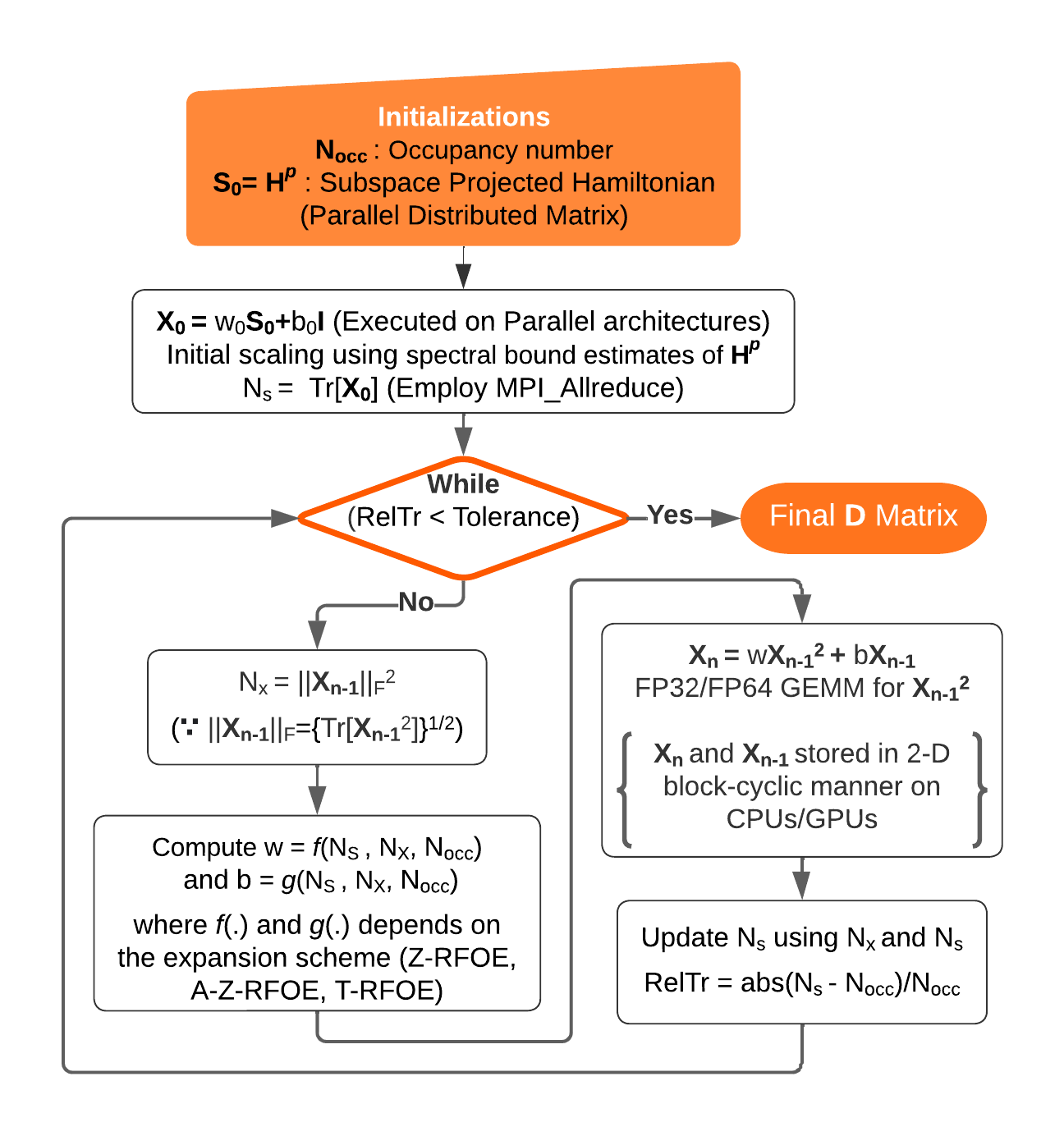}
    \caption{General implementation details flowchart for all the  RFOE codes \centering}
    \label{fig:RFOESchematic}
\end{figure}

Figure~\ref{fig:RFOESchematic} describes the overall algorithmic details of implementing recursive Fermi-operator expansion (RFOE) approaches discussed in this work (Z-RFOE, A-Z-RFOE and T-RFOE) employing $\Hmatp$. As indicated in the figure, the trace change in the trial density matrix ($\bX_n$) for a given $n^{th}$ iteration relative to the previous iteration ($\mathtt{RelTr}$) is used as a criterion for RFOE convergence. Furthermore, we also explore mixed-precision strategies in conjunction with the Z-RFOE and T-RFOE schemes implemented in this work. To this end, we rely on the fact that far away from RFOE convergence to the density matrix, the floating point operations can be performed in single precision (FP32) and switching to double precision (FP64) operations thereafter. The criteria to decide the number of initial FP32 iterations is linked to $\mathtt{RelTr}$ . 
\vspace{-0.2in}
\subsection{\label{sec:level11}Implementation strategies}
\vspace{-0.1in}
The C++ implementation of RFOE codes employs the Message Passing Interface (MPI) library to achieve multi-node parallelism. To handle the parallel matrices encountered during the RFOE process, we utilize the SLATE (Software for Linear Algebra Targeting Exascale) library \cite{SLATE}. SLATE effectively stores these matrices in a 2-D block-cyclic fashion on both CPUs and GPUs within a node. The tile size, which determines the size of the sub-matrices (blocks) distributed in a 2-D block-cyclic manner across the processor grid, plays a crucial role in the parallel performance of matrix-matrix multiplication. Therefore, we conduct numerical experiments to explore the impact of varying the tile size on the overall performance. These experiments aimed to identify the optimal tile size for a given matrix dimension $\Hmatp$ .

The computation of the matrix trace, required at the beginning and end of RFOE iterations, involves traversing the diagonal elements of the global matrix and utilizing MPI collectives. Additionally, we compute the trace of the squares of symmetric matrices arising during the course of RFOE iterations by evaluating the square of the Frobenius norm of the given parallel distributed SLATE matrix (${Tr(\bA^2)} = ||\bA||^2_F$). This is facilitated through a SLATE API call, which grants users access to the Frobenius norm. Among the various steps involved in RFOE algorithms, matrix-matrix multiplication (MM) is the computationally dominant step. To handle this step efficiently across multiple CPUs and GPUs, we leverage the capabilities of the SLATE library, which enables parallel execution of the underlying MM. Furthermore, we explored the performance of alternative approaches, namely the Communication-optimal Matrix Multiplication (COSMA) \cite{COSMA} and cuBLASMg (provided by the CUDA Math Library Early Access Program \cite{cuBLASMG}) libraries, for computing parallel matrix-matrix multiplications on CPUs and GPUs. Our studies indicate that COSMA proved to be slower compared to the SLATE library in terms of computational times. Additionally, it's worth noting that the cuBLASMg library is limited to utilizing multiple GPUs within a single node, which constrains its applicability.

\vspace{-0.2in}
\subsection{\label{sec:level14}Metrics for accuracy benchmarking of RFOE}
\vspace{-0.1in}
For accuracy benchmarking of the RFOE methods implemented in this work, we compute two error metrics -- (a) relative error ($\epsilon_1$) in the Frobenius norm between the exact density matrix ($\bD^{p}_{ref} = f(\Hmatp)$) and the approximate density matrix ($\bD^{p}$) computed -via- RFOE  i.e. $\epsilon_1 = (||\bD^{p}-\bD^{p}_{ref}||_F)/||\bD^{p}_{ref}||_F$, and (b) relative error ($\epsilon_2$) between the trace of actual and the approximate value of $f(\Hmatp)\Hmatp$, a measure of band energy error. To this end, we have $\epsilon_2 = (Tr(\bD\Hmatp)-Tr(\bD_{ref}\Hmatp))/Tr(\bD_{ref}\Hmatp)$. $\bD_{ref}$ is computed by explicit diagonalization using ELPA library \cite{ELPA,ELPA2}.  We note that $||\bD^{p}_{ref}||_F$ will be of the $O(N_{occ})$, and hence the electron-density computed using RFOE approaches will have an absolute error $\epsilon_{\rho} = O(\epsilon_1 N_{occ})$ with respect to the true electron-density (ground-state). Since the Kohn-Sham energy functional is variational with respect to the ground-state electron density, the ground-state energy error will be of $O(\epsilon_{\rho}^2)$ and forces which are derivatives of ground-state energy with respect to the position of atoms will have an error of $O(\epsilon_{\rho})$.
\vspace{-0.2in}
\section{\label{sec:level15}Results and Discussion}
\vspace{-0.1in}
To assess the accuracy and performance of the proposed methods, we employ synthetic matrices representative of the subspace projected Hamiltonians ($\Hmatp$) arising in pseudopotential DFT calculations. To this end, the matrix $\Hmatp$ is constructed so that the spectral width is smaller and remains constant with an increase in the matrix size (indicative of an increase in the system size of a given material system). We choose the matrix elements of $\Hmatp$ such that $H^{p}_{ij} = H^{p}_{ji} = e^{-0.5 * |i - j|} * \sin(i + 1)$, and the matrix sizes investigated in this study are $8192\times8192$, $16384\times16384$, $32768\times32768$, and $65536\times65536$. The number of occupied states ($N_{occ}$) is chosen to be around 90\% of the dimension of the matrix in the case of zero-temperature density matrix, while in the case of the finite-temperature density matrix, the number of occupied states (including fractionally occupied) is chosen to be 85\% of the dimension of the matrix.  We compare the performance of different implementations of RFOE (Z-RFOE, A-Z-RFOE, T-RFOE) and explicit diagonalization of $\Hmatp$ using the ELPA~\cite{ELPA,ELPA2} library employing two-stage diagonalization algorithm (CPU-ELPA2, GPU-ELPA2) on both multi-node CPUs and GPUs. 

The multi-node CPU study used the PARAM PRAVEGA supercomputer equipped with 584 nodes. Each node has 2 Intel Xeon Cascade-Lake 8268, 2.9 GHz processors with 48 physical cores per node. The nodes are integrated with a BullSequana XH200 Mellanox HDR 100Gbps Infiniband interconnection using FAT-Tree topology for MPI.  On PRAVEGA, we have compiled our codes using the libraries --- GCC/11.2.0, INTEL-MKL/2020, INTEL-MPI/2019.10.317.  The multi-node GPU study was done on SUMMIT (Oak Ridge Leadership Computing Facility) supercomputer equipped with 4608 IBM Power System AC922 nodes with two IBM POWER9 processors (42 physical cores) and six NVIDIA Volta V100 GPUs in each node. Summit nodes are connected to a dual-rail EDR InfiniBand network providing a node injection bandwidth of 23 GB/s. On SUMMIT, we have compiled our codes using the libraries NVIDIA CUDA/11.0.3, GCC/9.1.0, IBM Spectrum-MPI/10.4.0.3, and OPENBLAS/0.3.15-omp



Throughout our discussion of the results in this section, it is important to note that a relative trace tolerance $\mathtt{RelTr}$ of $5\times 10^{-8}$ is utilized as a convergence criterion for all double-precision (FP64) RFOE implementations described here (see Fig.1). As a result, the number of RFOE iterations (parallel matrix-matrix multiplications) is observed to be around 50 in the case of Z-RFOE and around 30 in the case of A-Z-RFOE. For the mixed precision strategy implemented in the case of Z-RFOE and T-RFOE, the value of $\mathtt{RelTr}$ is chosen to be around $5\times 10^{-4}$ for FP32 RFOE iterations, and subsequently, the RFOE iterations in FP64 precision are continued till a $\mathtt{RelTr}$ of $5\times 10^{-8}$ is reached. In the case of finite-temperature density matrix approximation using T-RFOE, the smearing value of $\beta$ corresponding to a temperature of 500K is chosen. We observe the number of T-RFOE iterations to be around 30. Furthermore, we note that the tolerance criteria $\mathtt{RelTr}$ mentioned above resulted in errors $\epsilon_1$ and $\epsilon_2$ (refer to sec IIIC) of approximately $O(10^{-10})$ and $O(10^{-9})$, respectively, in the double-precision implementation of Z-RFOE and A-Z-RFOE. In the case of mixed precision implementation employing Z-RFOE, the errors $\epsilon_1$ and $\epsilon_2$ are of the $O(10^{-7})$ and $O(10^{-9})$ respectively. In the case of T-RFOE, we remark that the error $\epsilon_1$ is of the $O(10^{-3})$ and $\epsilon_2$ is of $O(10^{-6})$. We note that the higher errors associated with the finite-temperature density matrix approximated by T-RFOE relative to the exact Fermi-Dirac density matrix are expected, as the T-RFOE method \cite{TRFOE2} can be considered to be an alternative approach to smearing the zero-temperature density matrix. This viewpoint renders T-RFOE useful for approximating the finite-temperature density matrix in DFT since it relies on energy differences to compute material properties.

We now list performance metrics used for our comparative study in this work:
\begin{itemize}
    \item CPU Node-hrs $\Rightarrow$ Execution time (in hours) $\times$ the number of nodes (chosen in the good scaling regime). It gives a measure of computational efficiency on CPUs.
    \item GPU-hrs $\Rightarrow$ Execution time (in hours) $\times$ the number of GPUs (chosen in the good scaling regime). It gives a measure of computational efficiency on GPUs.
    \item Minimum wall time $\Rightarrow$  Least possible time obtained using as many resources (CPUs/GPUs) as possible. It is a measure of the scaling efficiency of the implementation.
    \item Energy consumption $\Rightarrow$ Upper bound of power consumption in kWh for executing a given algorithm on CPUs/GPUs. Indicative of the dollar cost required for the calculations on the supercomputer. For the energy consumption calculation, we used the Thermal Design Power (TDP) ratings for both CPUs and GPUs.
\end{itemize}
\vspace{-0.2in}
\subsubsection{\label{sec:level16}Multi-node CPU comparisons} 
\begin{figure}[!btp]
\centering
\subfloat[(a)]{\includegraphics[width=0.8\linewidth]{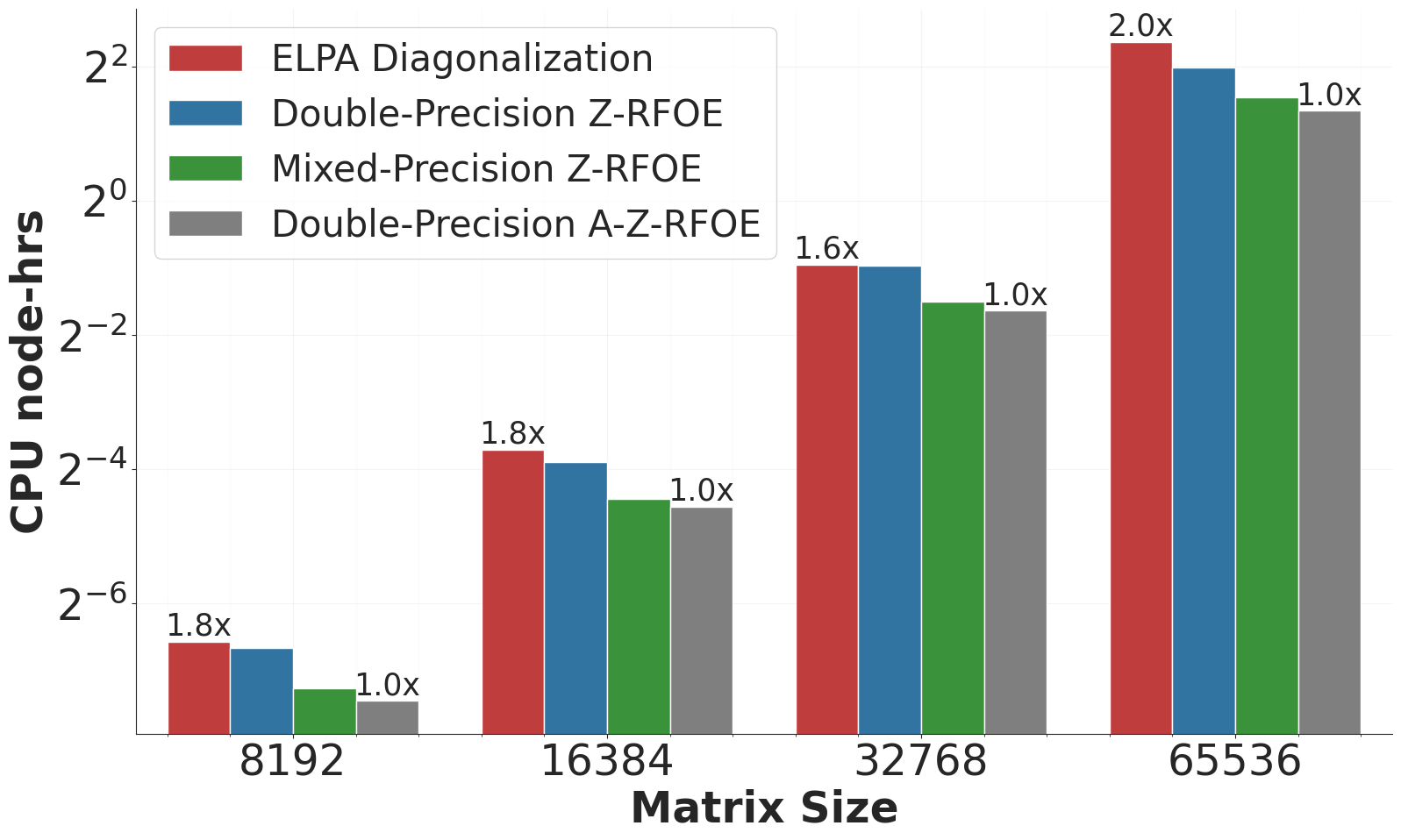}}
\hfill
\subfloat[(b)]{\includegraphics[width=0.8\linewidth]{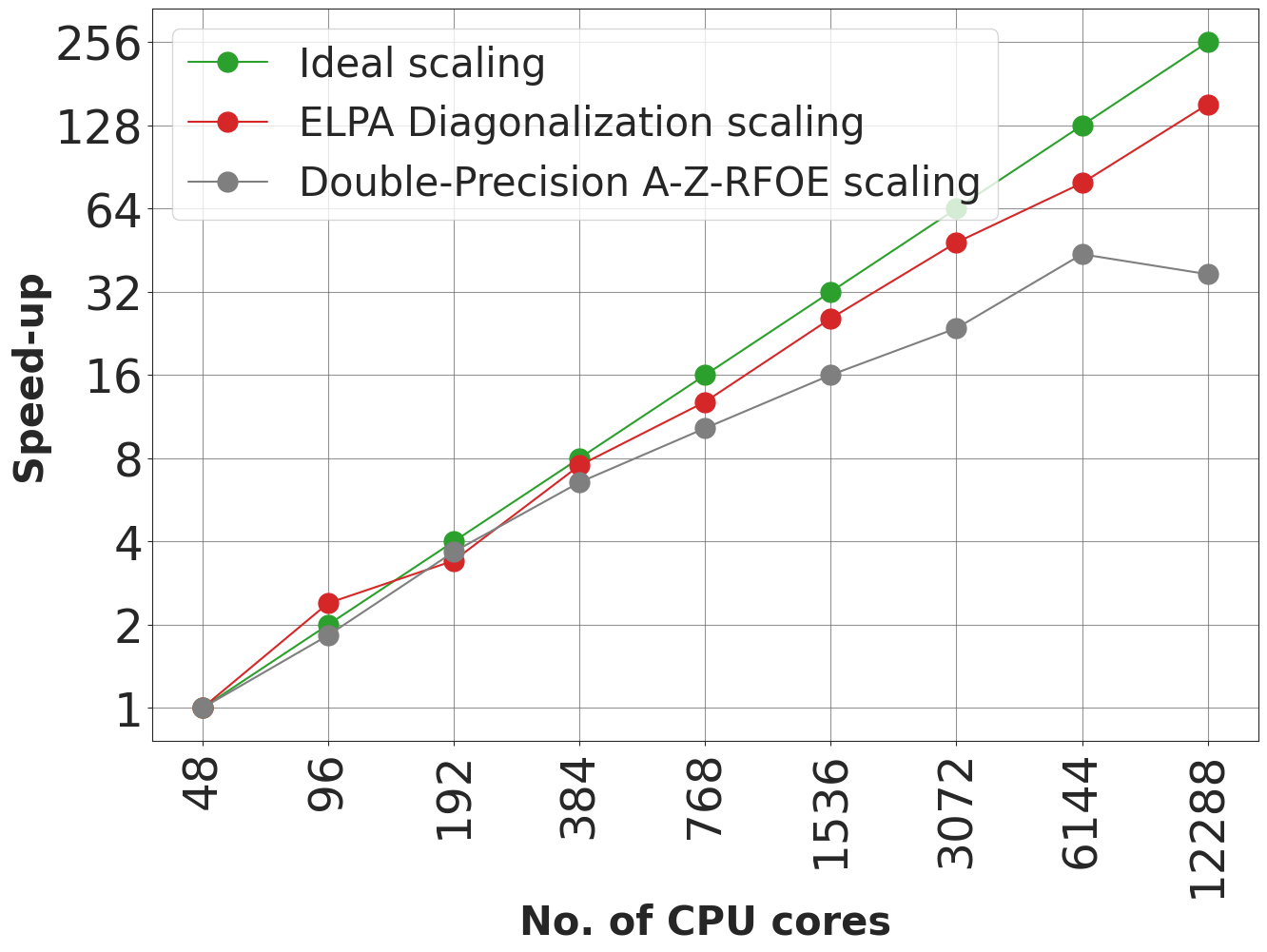}}
\hfill
\subfloat[(c)]{\includegraphics[width=0.8\linewidth]{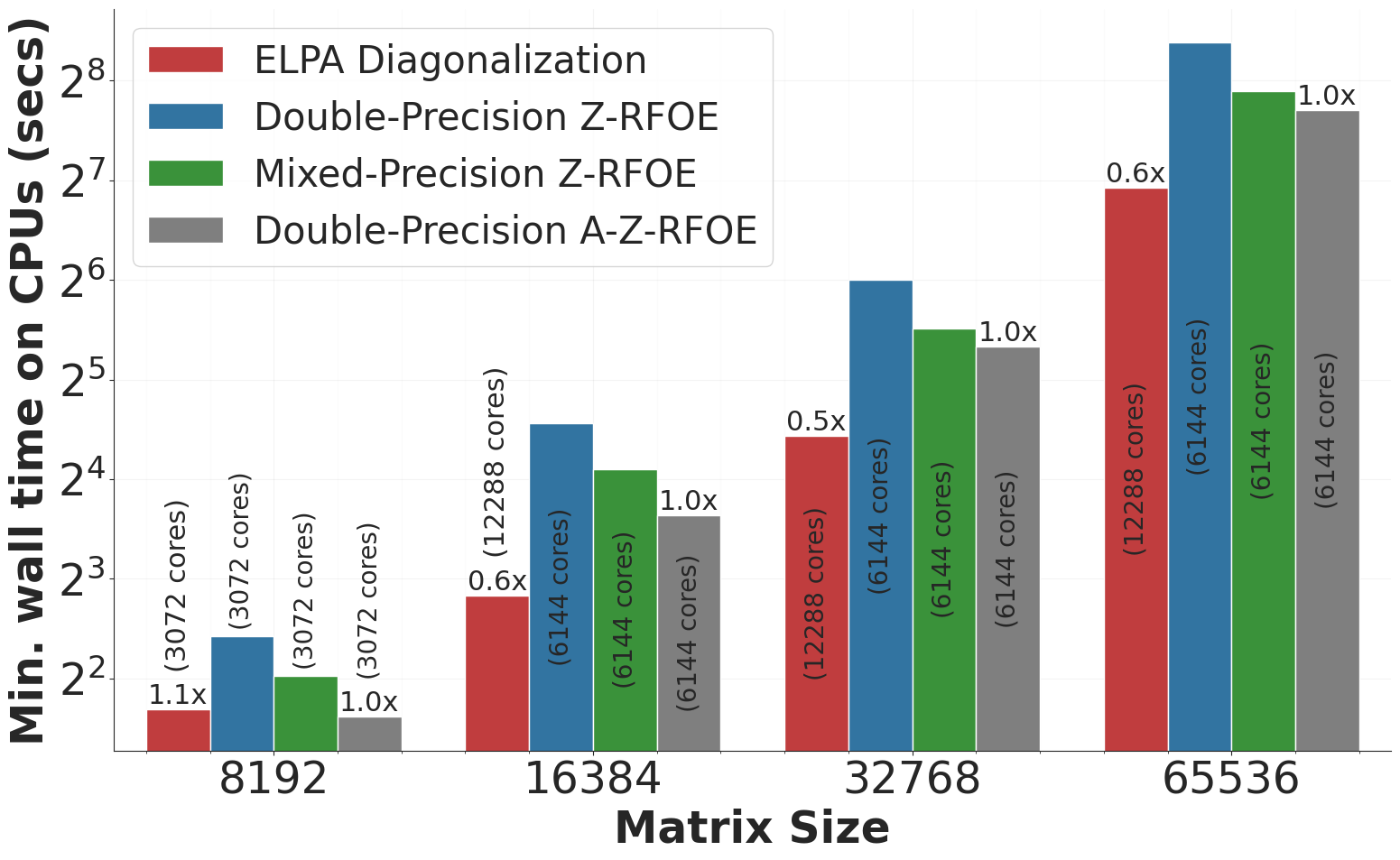}}
\caption{(a) CPU Node-hrs vs. matrix size plot, (b) Scaling study involving $\Hmatp$ of size 65536, and (c) Min. wall time vs. matrix size plot. Case study: ELPA diagonalization vs different implementations of RFOE used to approximate the zero-temperature density matrix on multi-node CPUs. The bar plots in (c) indicates the number of CPU cores at which minimum wall time is obtained for each matrix size. \raggedright}
    \label{fig:417}
\end{figure}

\begin{figure}[!btp]
\centering
\subfloat[(a)]{\includegraphics[width=0.8\linewidth]{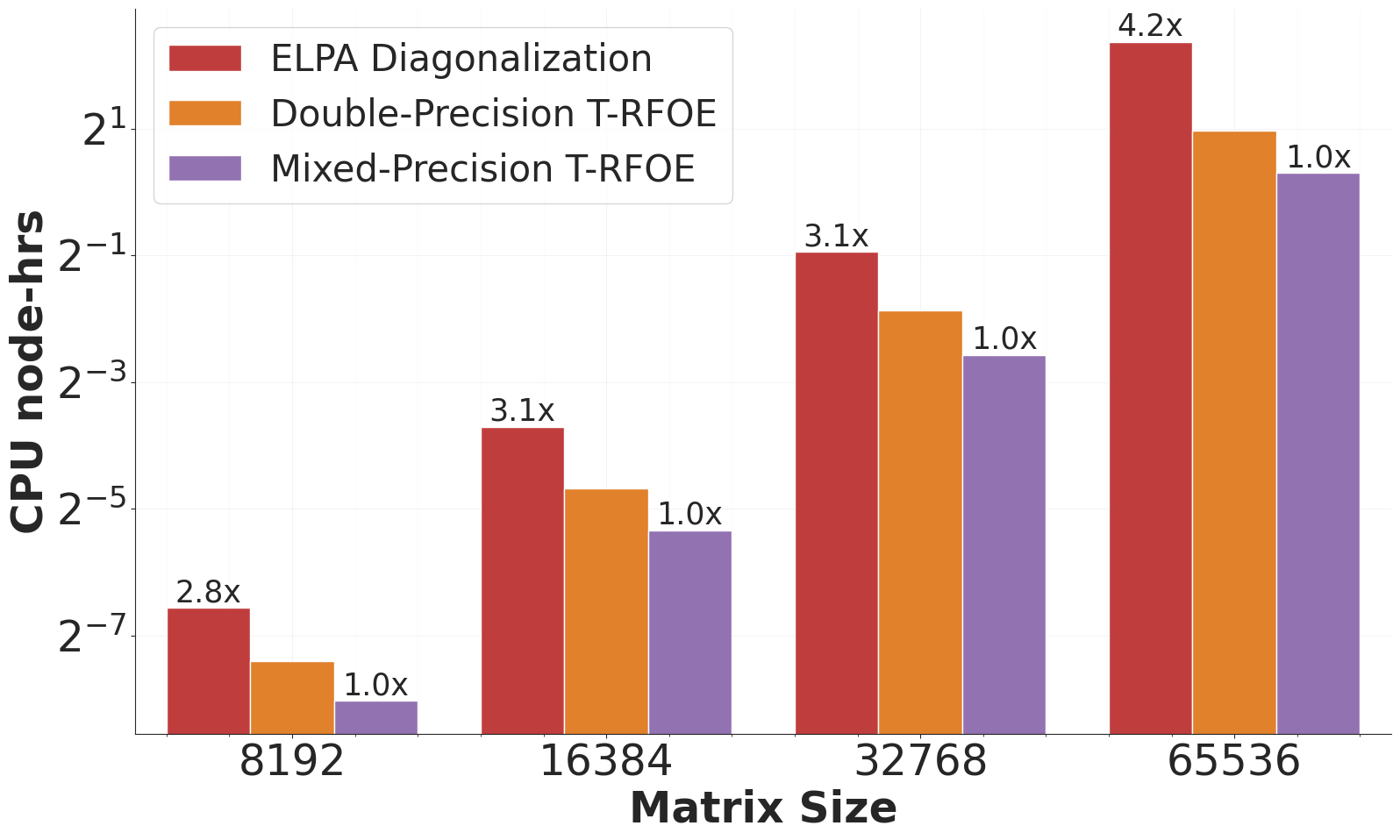}}
\hfill
\subfloat[(b)]{\includegraphics[width=0.8\linewidth]{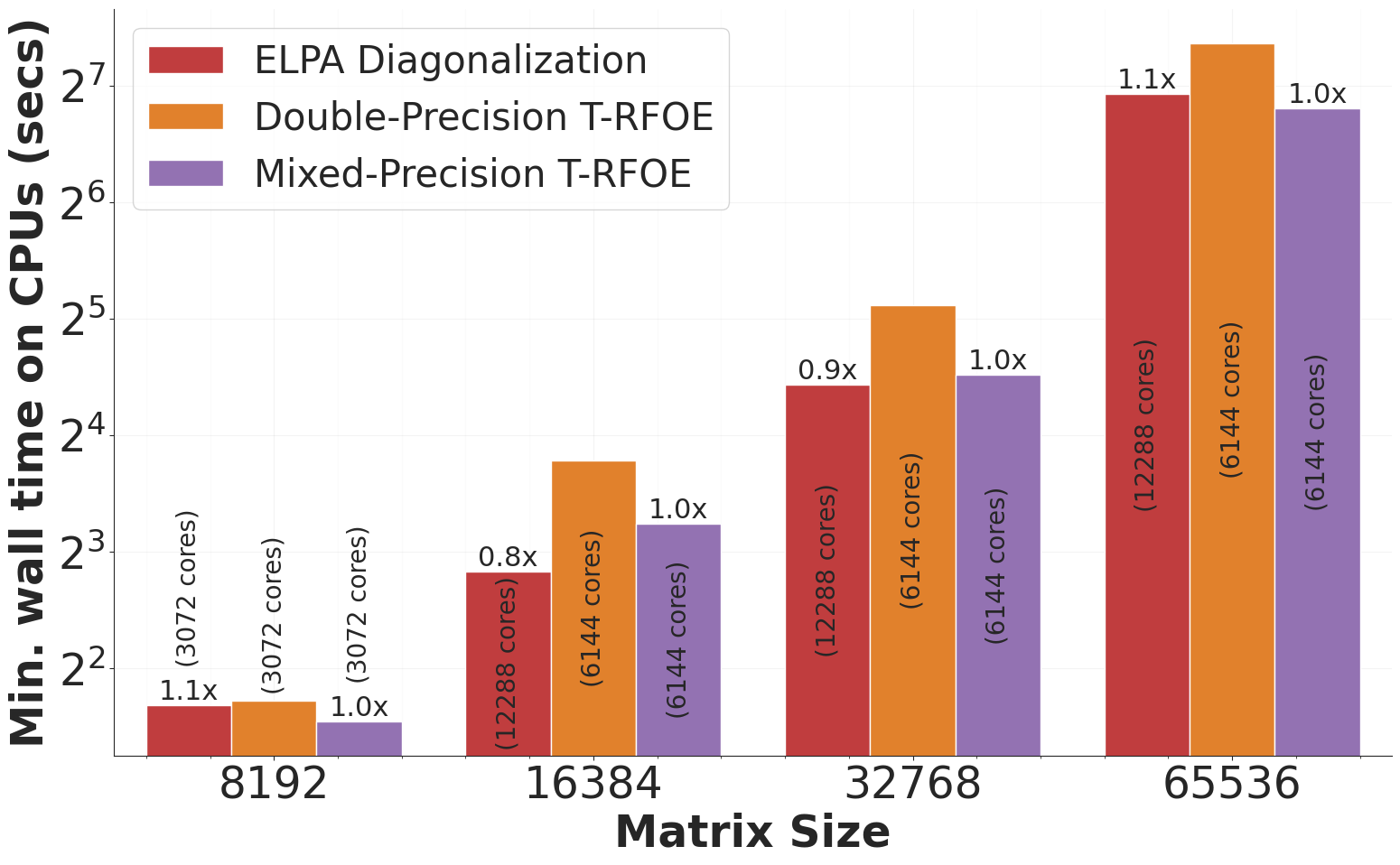}}
\caption{(a) CPU Node-hrs vs. matrix size plot, and (b) Min. wall time vs. matrix size plot for different implementations of RFOE used to approximate the finite-temperature density matrix on multi-node CPUs. The bar plots in (b) indicates the number of CPU cores at which minimum wall time is obtained for each matrix size \raggedright}
    \label{fig:418}
\end{figure}

\vspace{-0.1in}
Figures~\ref{fig:417} and~\ref{fig:418} provide a comparison between ELPA's subspace diagonalization procedure and various RFOE approaches that utilize parallel dense matrix-matrix multiplications based on the SLATE library. The comparison is based on CPU node-hrs, scalability, and minimum wall times on multi-node CPU architectures. On CPUs, Figure~\ref{fig:417}(a) demonstrates that RFOE implementations for approximating the zero-temperature density matrix are computationally efficient than subspace diagonalization in terms of CPU node-hrs. Among the RFOE implementations, A-Z-RFOE is the closest competing method to ELPA's diagonalization approach and is faster than ELPA approximately by a factor of 2x for all the matrix $\Hmatp$ dimensions (8192, 16384, 32768 and 65536) considered in this work. This can be attributed to the reduced computational prefactor in RFOE approach (only dense matrix-matrix multiplications) compared to exact diagonalization (reduction to tridiagonal matrix followed by reduction to Schur form). Further, in Figure \ref{fig:417}(b), the multi-node CPU scaling behavior of the subspace diagonalization procedure using ELPA and the A-Z-RFOE approach employing the SLATE library for the $\Hmatp$ matrix of dimension 65536 is illustrated. These results indicate the superior scaling behaviour of ELPA over SLATE. For instance, relative to 48 cores (1 node), the minimum number of nodes that the problem can be accommodated, we find the scaling efficiency of ELPA diagonalization is close to 75\% at 3072 cores (64 nodes) and drops to around 60\% at 12288 cores (256 nodes). In contrast, the scaling efficiency of A-Z-RFOE approach using SLATE drops to 37\% at 3072 cores (64 nodes) and flattens around 6144 cores (128 nodes). Insights into minimum wall times, i.e. the least possible execution time obtained with increasing CPU cores, are valuable outcomes of the scaling efficiency for various implementations considered in this work on multi-node CPU architectures\cn. To this end, Figure \ref{fig:417}(c) demonstrates that the minimum wall times obtained by ELPA's subspace diagonalization are faster by a factor of around 1.7x - 2x compared to A-Z-RFOE, the closest competitor and is consistent with the observation that ELPA's scaling behaviour is close to ideal scaling behaviour till large number of CPU cores compared to SLATE. For instance, in this case of the matrix $\Hmatp$ of dimension 65536, the minimum wall times are observed at around 12000 CPU cores when using an explicit diagonalization approach employing the ELPA library, while the best RFOE implementation (A-Z-RFOE) based on SLATE gave a 1.7x higher minimum wall time at around 6000 CPU cores.
Finally, Figure~\ref{fig:418} illustrates the comparisons for building the finite-temperature density matrix on multi-node CPUs using T-RFOE and diagonalization approaches. We observe from Figure~\ref{fig:418}(a) that T-RFOE approaches are more computationally efficient than ELPA's diagonalization in terms of CPU node-hrs. In particular, mixed-precision T-RFOE implementation has been observed to be faster by a factor of around 3x-4x compared to ELPA's diagonalization for the matrix dimensions considered in this work. And, in the minimum wall time comparisons illustrated in Figure~\ref{fig:418}(b), mixed precision T-RFOE is found to have almost similar wall times as that of diagonalization using ELPA for all the sizes of the matrices considered in this work. 

\captionsetup[subfigure]{labelformat=empty}

\captionsetup[subfigure]{labelformat=empty}
\vspace{-0.2in}
\subsubsection{\label{sec:level17}Multi-node GPU comparisons}
\vspace{-0.1in}

\begin{figure}[!btp]
\centering
\subfloat[(a)]{\includegraphics[width=0.8\linewidth]{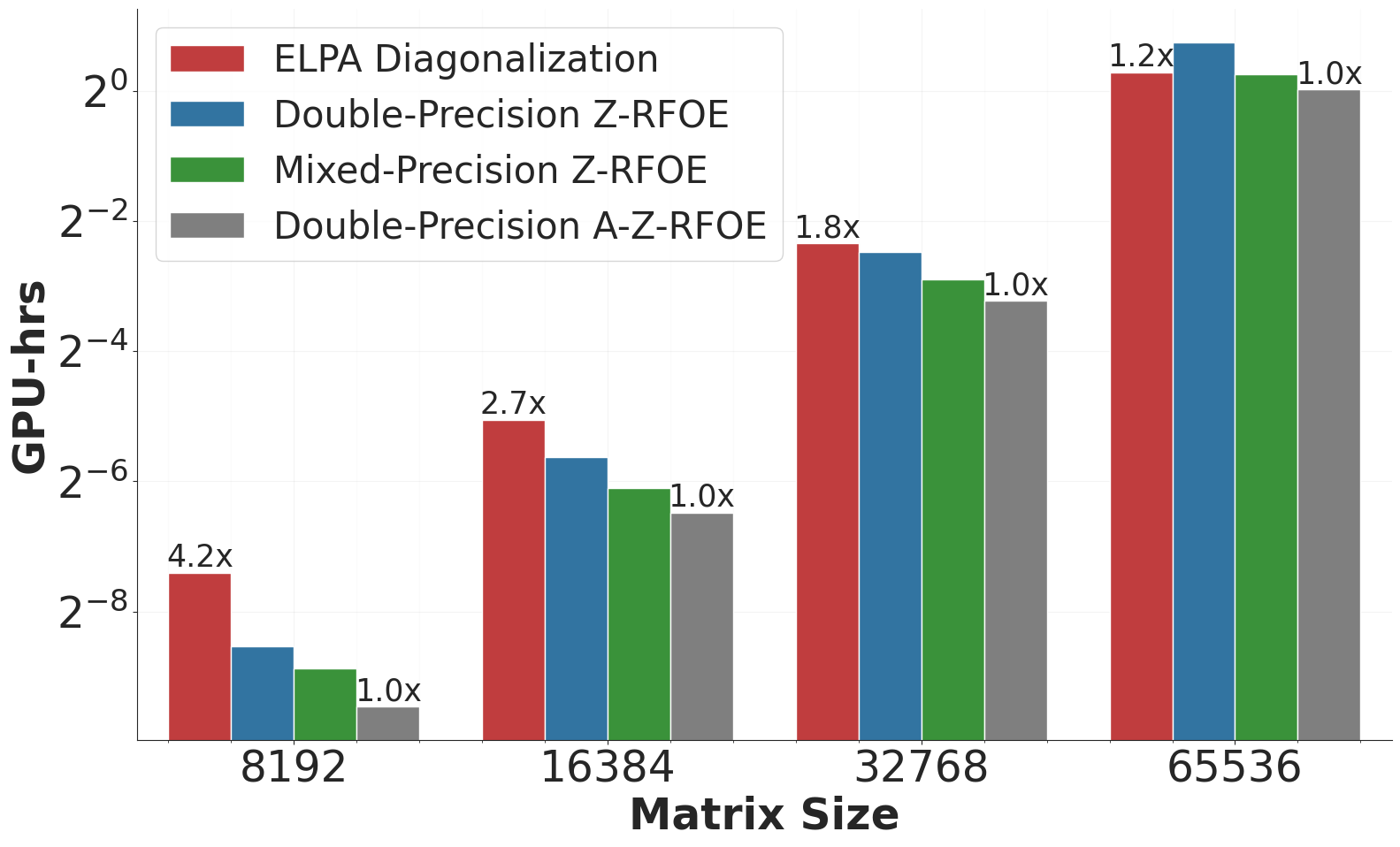}}
\hfill
\subfloat[(b)]{\includegraphics[width=0.8\linewidth]{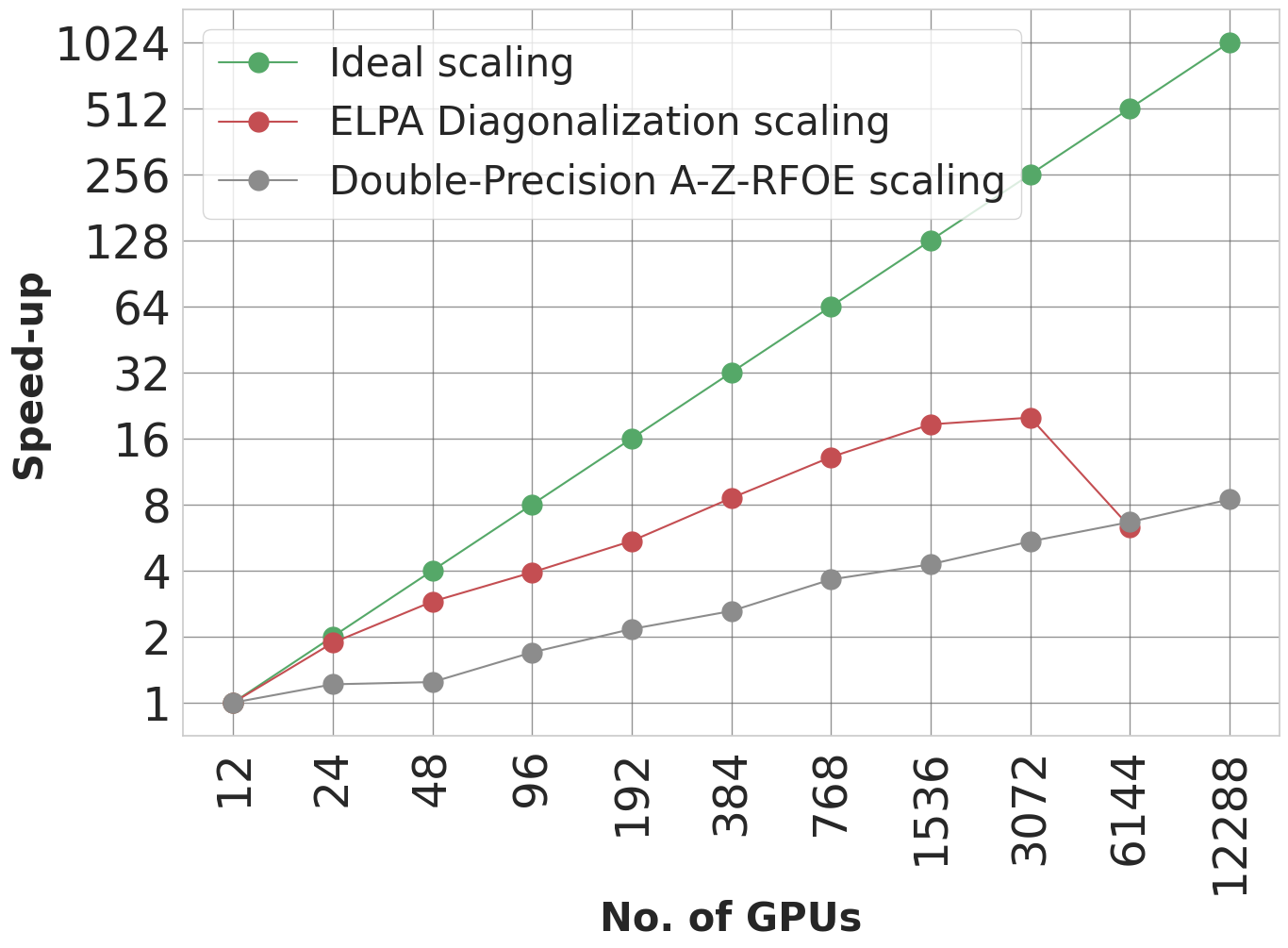}}
\hfill
\subfloat[(c)]{\includegraphics[width=0.8\linewidth]{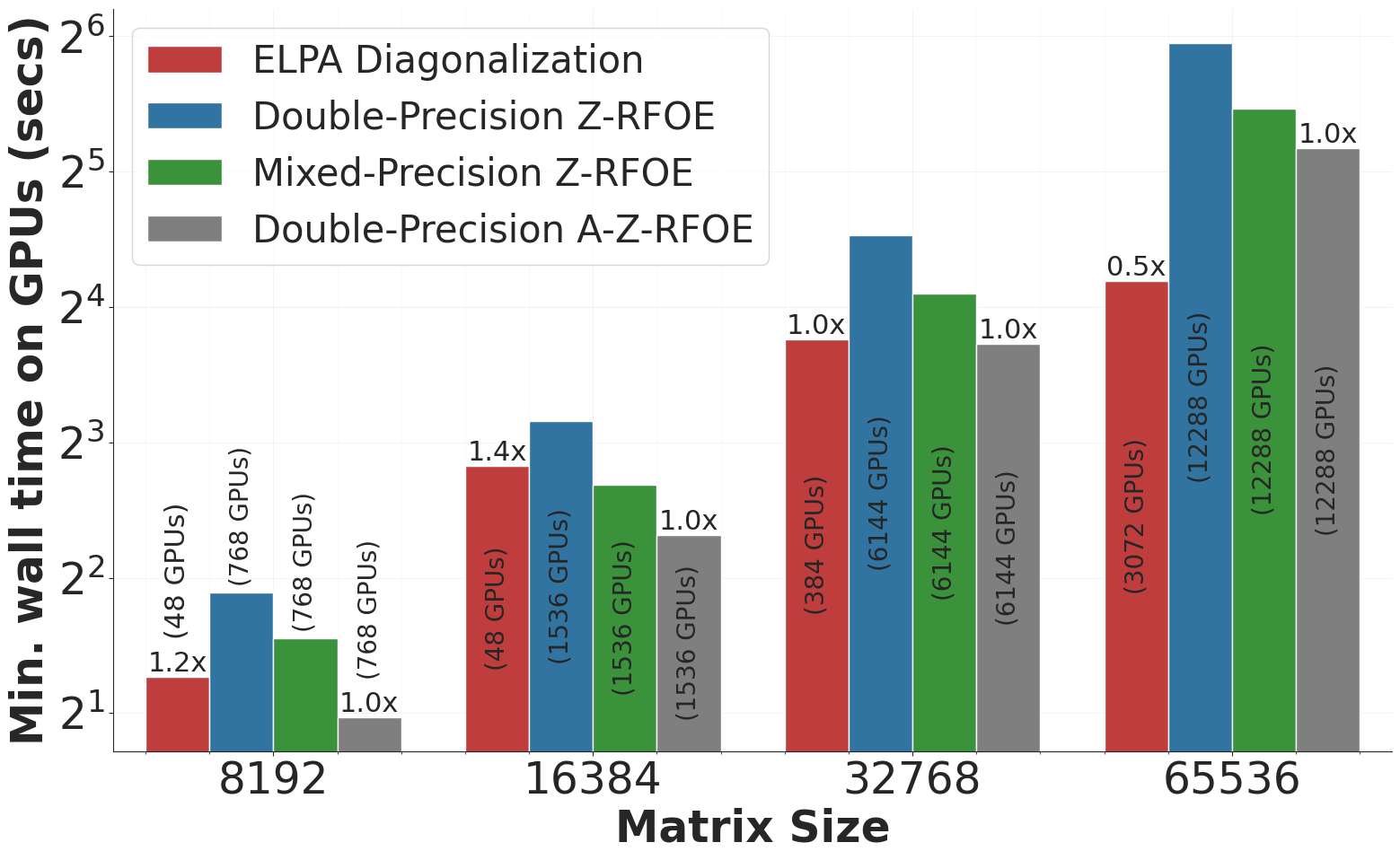}}
\caption{(a) GPU-hrs vs. matrix size plot, and (b) Scaling study (c) Min. wall time vs. matrix size plot. Case study: ELPA Diagonalization vs different implementations of RFOE used to approximate the zero-temperature density matrix on multi-node GPUs \centering}
    \label{fig:419}
\end{figure}
\begin{figure}[!btp]
\centering
\subfloat[(a)]{\includegraphics[width=0.8\linewidth]{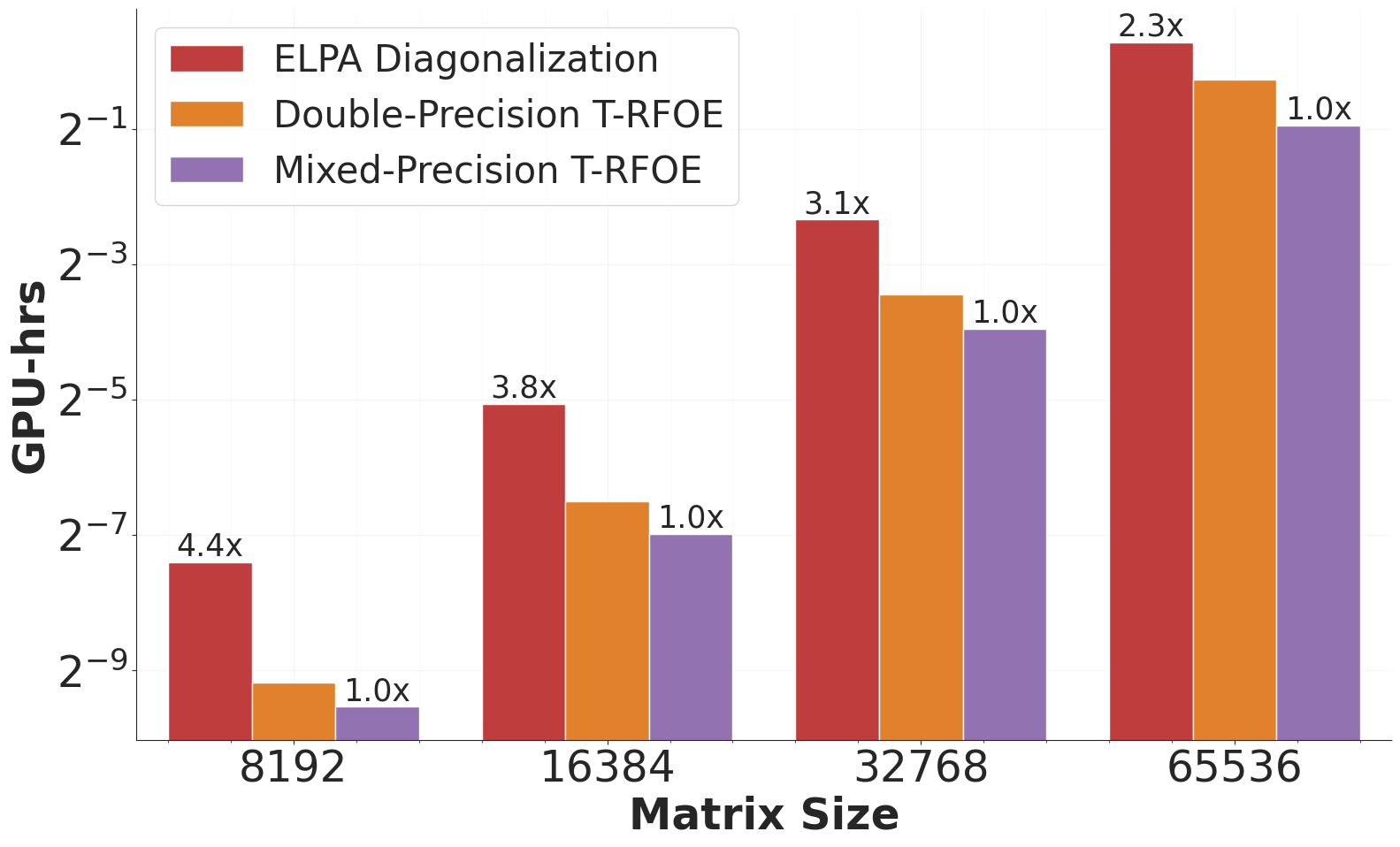}}
\hfill
\subfloat[(b)]{\includegraphics[width=0.8\linewidth]{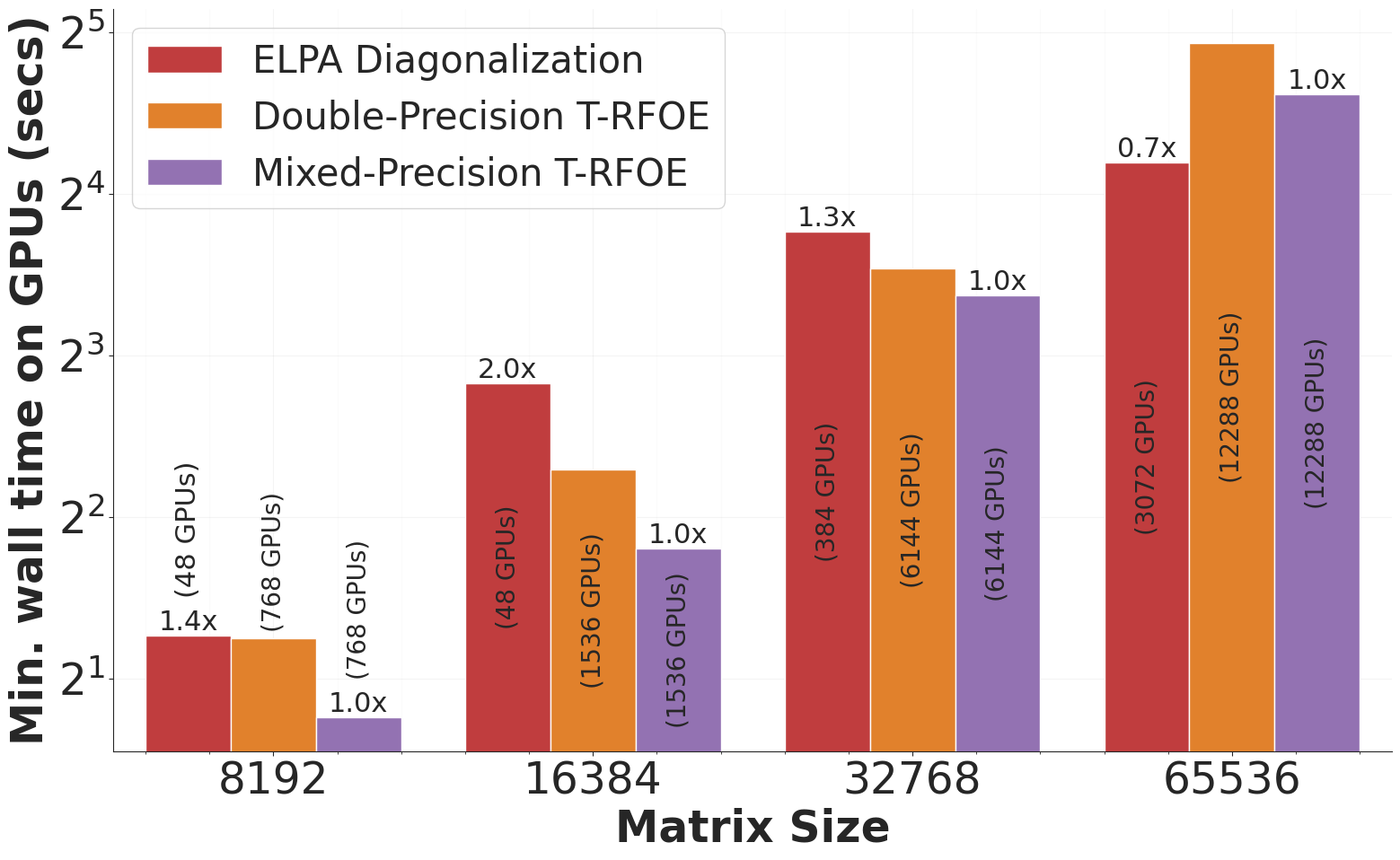}}
\caption{(a) GPU-hrs vs. matrix size plot, and (b) Min. wall time vs. matrix size plot for different implementations of RFOE used to approximate the finite-temperature density matrix on multi-node GPUs.\centering}
    \label{fig:420}
\end{figure}

Figures~\ref{fig:419} and~\ref{fig:420} illustrate a comparative study of GPU-hrs, scalability, minimum wall times on multi-node GPU architectures between ELPA's subspace diagonalization procedure and different RFOE approaches implemented using SLATE. On V100 GPUs, Figure~\ref{fig:419}(a) demonstrates that utilizing RFOE implementations for approximating the zero-temperature density matrix is computationally more efficient than subspace diagonalization in terms of GPU-hrs. Among the RFOE implementations, A-Z-RFOE is the closest competing method to ELPA's diagonalization procedure and is faster than ELPA approximately by a factor of 4x, 2.7x, 2x and 1.2x for matrix $\Hmatp$ dimensions of 8192, 16384, 32768 and 65536 respectively. These results indicate that the speed-up over ELPA reduces with an increase in the size of the matrix. We note that the total computation time observed in RFOE or ELPA diagonalization is the cumulative of data movement costs and floating point operations costs on GPUs. The reduction in speed-ups of RFOE over ELPA with increase in matrix size can be attributed to the reduction in the fraction of data movement cost to the total computational cost associated with ELPA's two phase diagonalization approach. Furthermore, Figure~\ref{fig:419}(b) shows the scalability data for both subspace diagonalization procedure using ELPA and A-Z-RFOE implemented using SLATE for $\Hmatp$ matrix of dimension 65536. These results indicate the superior scaling behaviour of ELPA over SLATE. For instance, relative to 2 nodes (12 GPUs), the minimum number of nodes that the problem can be accommodated, we find that the scaling efficiency of RFOE approaches using SLATE quickly drops to 60\% at 4 nodes (24 GPUs), 20\% at 16 nodes (96 GPUs), 10\% at 128 nodes (768 GPUs) and around 1\% at 2048 nodes (12288 GPUs). In contrast, the scaling efficiency for ELPA is around 93\%, 50\%, 20\% and 14\% at 4 nodes, 16 nodes, 128 nodes and 256 nodes, respectively. As discussed previously in the CPU comparisons, a practically valuable outcome of the scaling efficiency is the minimum wall time, i.e., the least possible execution time one can obtain with increasing GPUs. Figure~\ref{fig:419}(c) shows these minimum wall time comparisons between various approaches considered in this work. The minimum wall times obtained using A-Z-RFOE is faster by a factor of 1.2x - 1.4x compared to ELPA's subspace diagonalization for the $\Hmatp$ matrix of dimensions 8192, 16384. However, the superior scaling of ELPA over SLATE resulted in these minimum wall times on fewer GPUs using ELPA's diagonalization (48 GPUs) compared to RFOE approaches considered in this work (768 and 1536 GPUs for matrix dimensions of 8192 and 16384 respectively). As the matrix dimension increases to 32768, the minimum wall time obtained is almost similar in the case of diagonalization and A-Z-RFOE implementation using SLATE, while the number of GPUs to achieve this wall time is larger by a factor of 16x for A-Z-RFOE approach in comparison to diagonalization. In the case of 65536 matrix size, a factor of 2x speed-up is observed in minimum wall time employing ELPA's diagonalization with much fewer GPUs than A-Z-RFOE. Finally, Figure \ref{fig:420} illustrates the comparative studies carried out in building the finite-temperature density matrix on GPUs using T-RFOE and diagonalization approaches. We observe that both double and mixed-precision variants of the T-RFOE implementations are more computationally efficient than the explicit diagonalization procedure using ELPA in the GPU-hrs regime. We also observe that mixed-precision variant of T-RFOE is the closest competitor to the diagonalization approach. In particular, we see computational gains of around 4.4x, 3.8x, 3x and 2.3x in terms of GPU-hrs for the $\Hmatp$ matrix of dimensions 8192, 16384, 32768 and 65536, respectively, over ELPA's diagonalization. Figure \ref{fig:420}(b) illustrates the minimum wall time comparisons, and these times obtained using mixed-precision T-RFOE are faster by a factor of 1.3x - 2x compared to ELPA's subspace diagonalization involving matrix dimensions of 8192, 16384 and 32768. However, the superior scaling of ELPA over SLATE resulted in these minimum wall times on fewer GPUs using ELPA. Further, for the $\Hmatp$ matrix size of 65536, the minimum wall time obtained -via- ELPA's diagonalization approach was lower than T-RFOE and was obtained on much fewer GPUs. \cn 

\cb The comparisons reported in this study were performed on V100 SXM2 GPUs. However, it is important to note that newer GPU architectures can potentially provide even greater computational improvements in terms of GPU-hrs. This is because the dominant computational costs associated with the RFOE approaches examined in this study are primarily associated with dense matrix-matrix multiplications. For instance,  we anticipate a speed-up of $\smallsim 1.5-2\times$ in terms of GPU-hrs for RFOE approaches using A100 SXM2 GPUs in comparison to V100 SXM2, taking into account the data movement costs associated with dense matrix-matrix multiplications. Futhermore, for the case of H100 GPUs, we anticipate the speed-ups in GPU-hrs compared to V100 GPUs can be around $\smallsim 5-6\times$ since H100 GPUs have close to 8$\times$ higher FP64 throughput performance. These advanced architectures also can benefit ELPA's diagonalization but not to the extent it can benefit RFOE dense matrix-matrix multiplications since explicit diagonalization methods usually do not involve high arithmetic intensity operations like RFOE. This is because diagonalization relies on a two-phase approach with the first phase involving the reduction of a dense matrix to a tridiagonal matrix using Householder reflectors which are constructed using vectors and do not involve arithmetic intensity operations like dense matrix-matrix multiplications. The subsequent phase reduces this tridiagonal matrix to a diagonal matrix which involves manipulation with sparse matrices which involves more data movement costs. 

In the case of minimum wall time comparisons with the newer architecture GPUs, the speed-ups gained by RFOE approach over ELPA or vice versa may not change much compared to what is reported in the manuscript as the communication cost becomes the dominant cost than compute at a larger number of GPUs. As of today, this is better handled in ELPA's diagonalization implementation than the parallel dense matrix-matrix multiplications (pgemm) implemented in the state-of-the-art libraries. \cn


\captionsetup[subfigure]{labelformat=empty}

\captionsetup[subfigure]{labelformat=empty}
\vspace{-0.2in}
\subsubsection{\label{sec:level18}Energy consumption comparisons}
\begin{figure}[!btp]
\centering
\subfloat[(a)]{\includegraphics[width=0.8\linewidth]{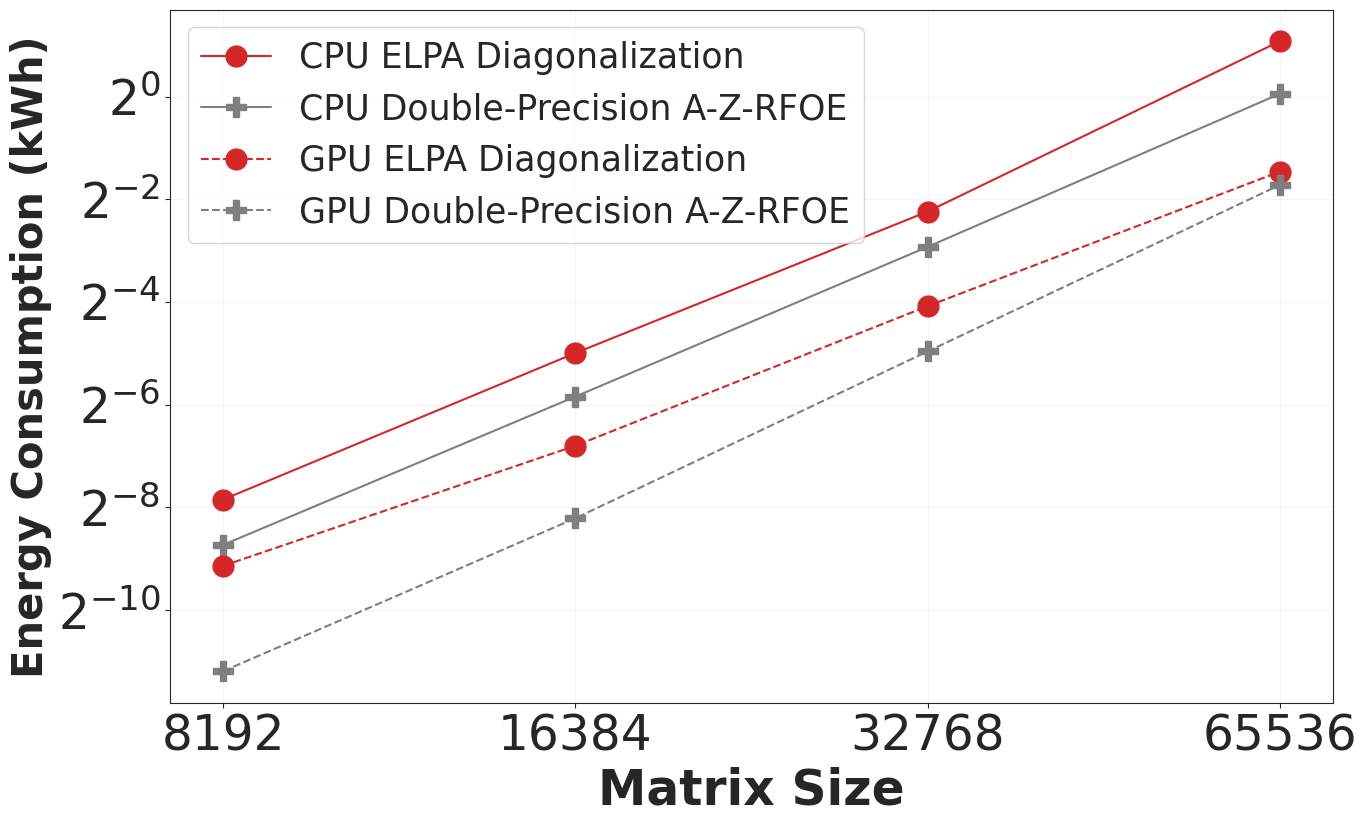}}
\hfill
\subfloat[(b)]{\includegraphics[width=0.8\linewidth]{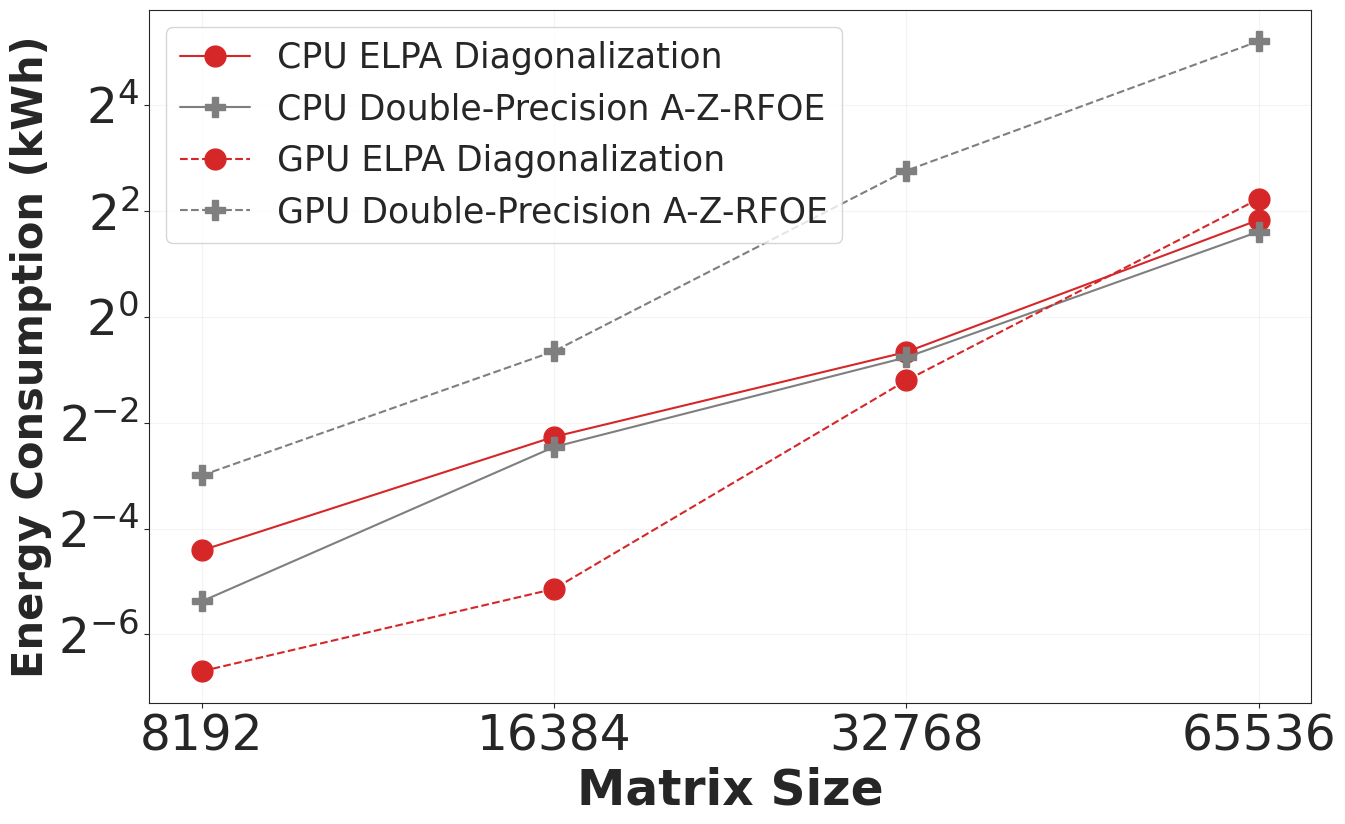}}
\caption{Energy consumption (kWh) vs. matrix size plot in terms of (a) Node-hrs/GPU-hrs, and (b) Min. wall time for the best-performing implementation of RFOE used to approximate zero-temperature density matrix.\centering}
\label{fig:421}
\end{figure}

\begin{figure}[!btp]
\centering
\subfloat[(a)]{\includegraphics[width=0.8\linewidth]{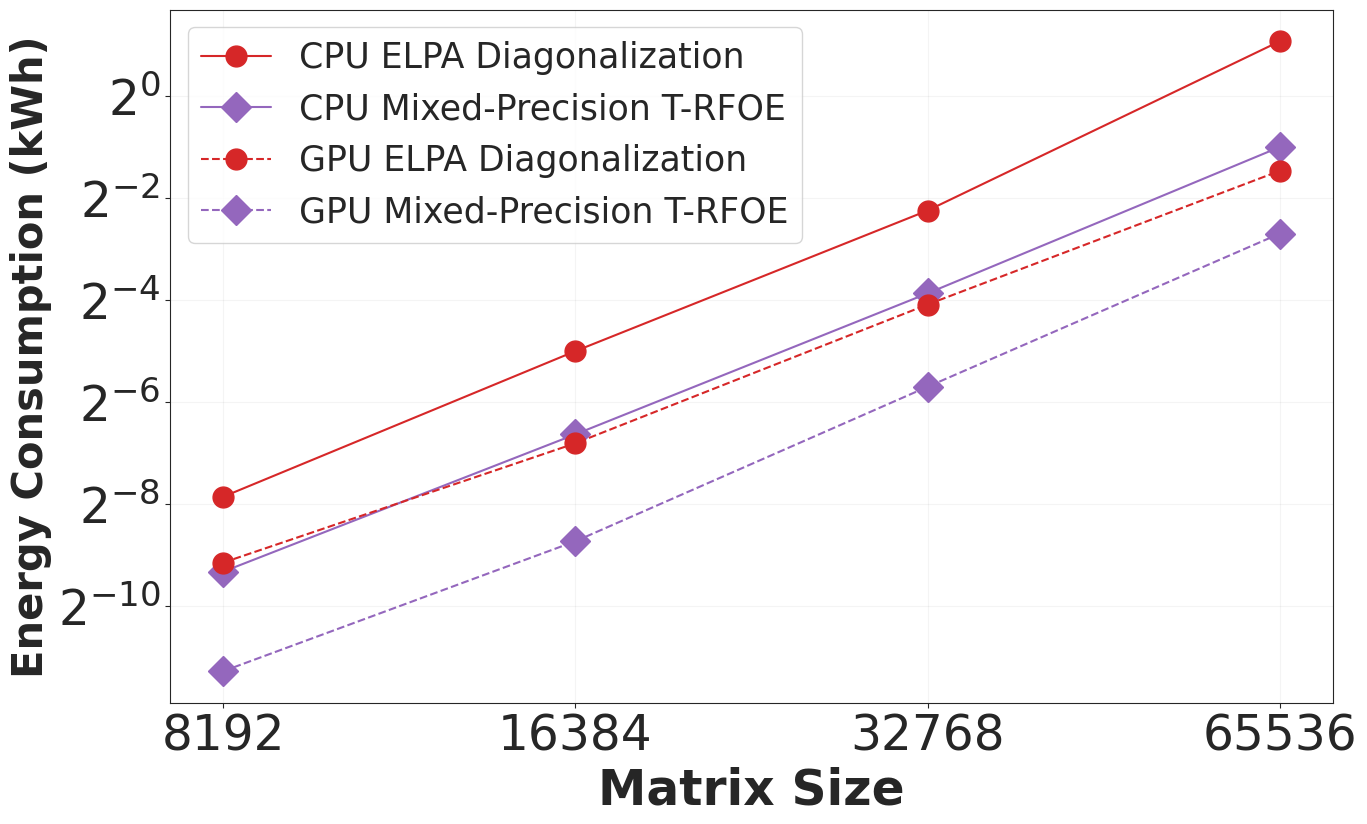}}
\hfill
\subfloat[(b)]{\includegraphics[width=0.8\linewidth]{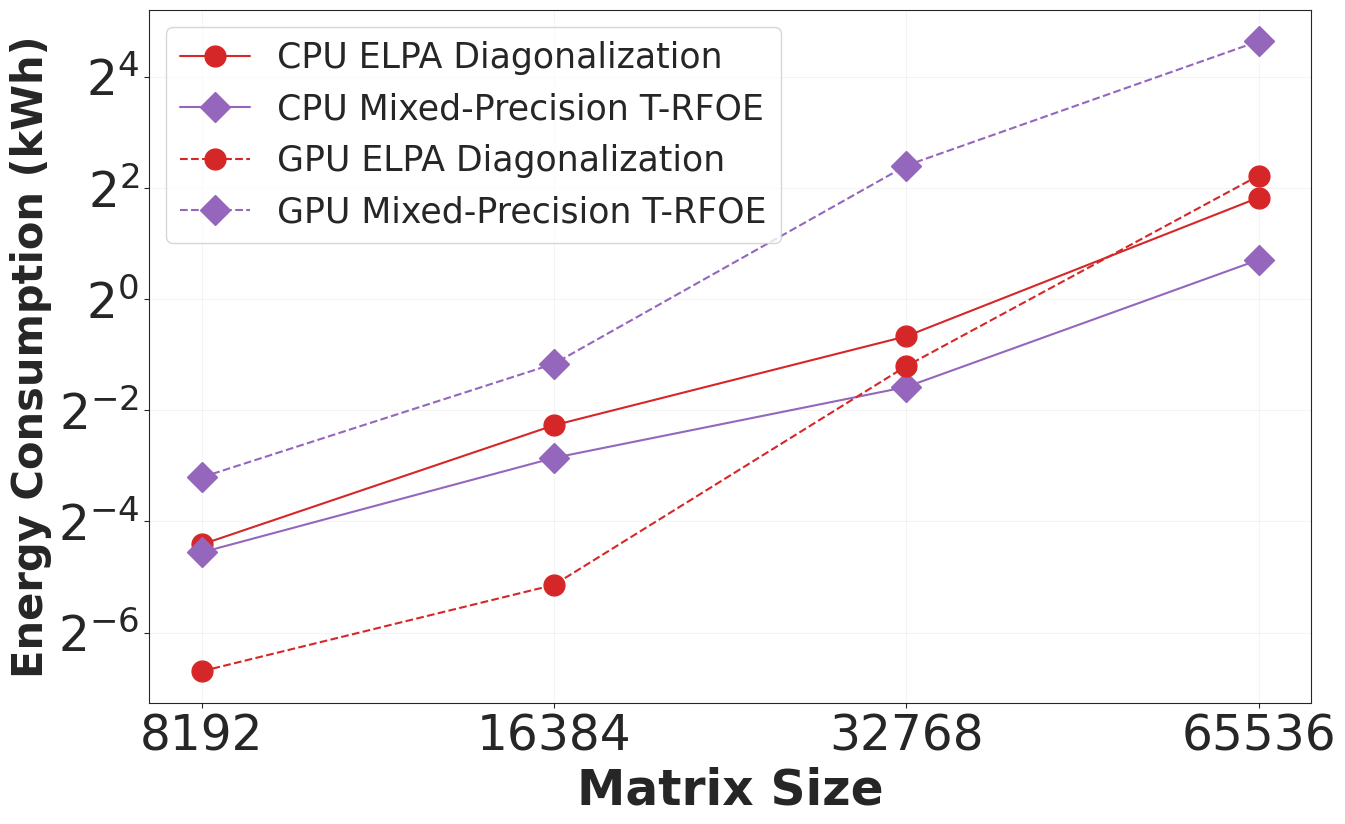}}
\caption{Energy consumption (kWh) vs. matrix size plot in terms of (a) Node-hrs/GPU-hrs, and (b) Min. wall time for the best-performing implementation of RFOE in the case of approximating finite-temperature density matrix.\centering}
    \label{fig:422}
\end{figure}

\vspace{-0.1in}
In order to estimate the upper bound for energy consumption in units of kWh as a function of matrix dimension, we utilize the TDP(Thermal design power) ratings for executing both RFOE implementations and the explicit diagonalization approach using the ELPA library. To plot the energy consumption against the matrix dimensions, we consider both the CPU node-hours/GPU-hours and minimum wall time regimes using the execution time of the best-performing RFOE implementation and ELPA diagonalization involving $\Hmatp$. Figures \ref{fig:421} and \ref{fig:422}summarize the relevant data. As the matrix size increases, we observe a higher energy consumption in the case of explicit diagonalization using ELPA for CPUs, both in the regime of CPU node-hours ($\smallsim 2\times$) and minimum wall time ($\smallsim 1.2\times$) compared to the best-performing RFOE implementations. 
In the case of GPUs, we also observe a higher energy consumption for explicit diagonalization using ELPA in the regime of GPU-hrs. However, in the regime of minimum wall time, energy consumption is significantly higher in the case of best RFOE implementation using SLATE compared to ELPA, owing to the excellent scaling behaviour of ELPA on multinode GPUs.

\captionsetup[subfigure]{labelformat=empty}

\captionsetup[subfigure]{labelformat=empty}

\vspace{-0.2in}
\section{\label{sec:level20}Conclusions}
Our current work investigates recursive Fermi-operator expansion (RFOE) strategies involving subspace projected Hamiltonian as an alternative to explicit subspace diagonalization in iterative orthogonal projection methods for solving large-scale eigenvalue problems. We explore the potential use of these strategies on both multi-node CPU and GPU architectures by employing ELPA and SLATE libraries. As demonstrated by various studies conducted in this work, we find that RFOE schemes involving dense matrix-matrix multiplications have a lesser computational prefactor associated with the cubic scaling complexity and hence, are more computationally efficient ($\smallsim 2\times -\,4\times$) than the explicit diagonalization of the subspace projected Hamiltonian in the CPU node-hrs/GPU-hrs regime. Further, we found that RFOE strategies can be memory efficient compared to explicit diagonalization procedures on GPUs. However, our results also demonstrate that the scaling efficiency associated with the explicit subspace diagonalization procedure carried out using ELPA is superior to RFOE strategies resulting in lower minimum wall times on both CPUs and GPUs. Finally, we conclude that RFOE strategies can be considered as effective alternatives to subspace diagonalization, particularly in the context of the good scaling regime, where computational efficiency is measured in terms of CPU node-hrs/GPU-hrs consumed. However, it's worth noting that these strategies may not perform as well in the extreme scaling regime as demonstrated in the current work.
\vspace{-0.2in}
\section*{Acknowledgments}
The authors gratefully acknowledge the seed grant from Indian Institute of Science (IISc) and SERB Startup Research Grant from the Department of Science and Technology India (Grant Number:SRG/2020/002194) for the purchase of a GPU cluster, which provided computational resources for this work. The research used the resources of PARAM Pravega at Indian Institute of Science, supported by National Super-computing Mission (NSM) R\&D for exa-scale grant (DST/NSM/R\&D\_Exascale/2021/14.02). This research also used resources of the Oak Ridge Leadership Computing Facility at the Oak Ridge National Laboratory, which is supported by the Office of Science of the U.S. Department of Energy under Contract No. DE-AC05-00OR22725.

\nocite{*}
\bibliography{aipsamp}

\end{document}